# Temperature-dependent magnetic properties of a magnetoactive elastomer: immobilization of the soft-magnetic filler

Andrii V. Bodnaruk[1], Alexander Brunhuber[2], Viktor M. Kalita[1,3], Mykola M. Kulyk[1], Andrei A. Snarskii[3,4], Albert F. Lozenko[1], Sergey M. Ryabchenko[1], Mikhail Shamonin[2*]

[1]*Institute of Physics NAS of Ukraine, Prospekt Nauky 46, 03028 Kiev, Ukraine*
[2]*East Bavarian Centre for Intelligent Materials (EBACIM), Ostbayerische Technische Hochschule Regensburg, Prüfeninger Strasse 58, 93049 Regensburg, Germany*
[3] *National Technical University of Ukraine "Igor Sikorsky Kyiv Polytechnic Institute", Prospekt Peremohy 37, 03056 Kiev, Ukraine*
[4]*Institute for Information Recording NAS of Ukraine, Shpaka Street 2, 03113 Kiev, Ukraine*

Magnetic properties of a magnetoactive elastomer (MAE) filled with µm-sized soft-magnetic iron particles have been experimentally studied in the temperature range between 150 K and 310 K. By changing the temperature, the elastic modulus of the elastomer matrix was modified and it was possible to obtain magnetization curves for an invariable arrangement of particles in the sample as well as in the case when the particles were able to change their position within the MAE under the influence of magnetic forces. At low (less than 220 K) temperatures, when the matrix becomes rigid, the magnetization of the MAE does not show a hysteresis behavior and it is characterized by a negative value of the Rayleigh constant. At room temperature, when the polymer matrix is compliant, a magnetic hysteresis exists and exhibits local maxima of the field dependence of the differential magnetic susceptibility. The appearance of these maxima is explained by the elastic resistance of the matrix to the displacement of particles under the action of magnetic forces.

[*] Corresponding author. E-Mail: mikhail.chamonine@oth-regensburg.de

## 1. Introduction

In the present paper, we study magnetic properties of magnetoactive elastomers (MAEs) comprising soft magnetic (micrometer-sized, ferromagnetic) particles. By its structure, MAE is a composite material, where magnetic particles are separated by a non-magnetic elastomer matrix. The main feature of MAEs, which distinguishes them from conventional composite material, is that the matrix can be compliant and magnetized particles can change their position (displace and rotate) within an MAE under the action of magnetic forces [1-5]. The primary (methodological) goal of this article is to demonstrate an experimental method for studying the effect of magnetic-field-induced variations of the mutual arrangement of filler particles on the magnetic properties of an MAE. All material characteristics of the composite material in the present paper are effective parameters. This means that they refer to the entire sample as if the material would be replaced by a homogeneous medium with the same properties.

Investigation of MAEs is currently of great interest [6-11]. They revealed a giant magnetostriction [1,3,13-18], an anomalous magnetic-field dependence for the effective dynamic shear modulus [3,9,10,19,20], and an anomalous magnetic-field dependence of the effective dielectric constant and electrical conductivity [21-30]. These phenomena are unusual for classical magnetism. For example, the intrinsic magnetostriction of filler particles is by many orders of magnitude smaller than the magnetostriction observed in experiments on magnetized MAEs.

Several experiments on magnetization of MAEs have been already carried out in the past [1,31-34]. For comparison, the powders of ferromagnetic filler particles [1,31] as well as their suspensions [31] have been considered. A magnetic hysteresis has been reported in all these cases, even for soft magnetic inclusions where the intrinsic magnetic hysteresis is expected to be negligible. As far as MAEs are concerned, the magnetic hysteresis is known to be more pronounced for softer elastomer matrices if the concentration of particles remains constant [1,33]. The hysteresis is therefore commonly associated with the mobility of filler particles inside the polymer matrix [1,33] and the corresponding assemblage of magnetized particles into linear chain-like aggregates during the ascending magnetization branch and disintegration of these "chains" under the action of the elastic forces when the magnetic field is decreased [32]. Calculation of effective magnetic properties of highly filled composite materials is a complicated problem, which at present has no general solution [35]. In some approximations, theoretical considerations allow one to calculate the concentration and field dependences of the magnetic properties for composite materials with fixed ferromagnetic inclusions [36].

Note that previous experiments have been performed at room temperature and the derived conclusions referred to different samples. We propose a direct experiment, where for one and the same sample it is possible to obtain magnetization curves for the cases when the inclusions can change their positions and when the inclusions do not have such a possibility, remaining immobilized when the magnetic field $H$ is applied. The fact that discussed anomalous changes of magnetorheological and electric properties of MAEs depend on the elasticity of the matrix, is known from experiments on MAEs with matrices of different chemical composition [1,23,37,38]. We will use the fact that the rigidity of the elastomer of the matrix can be substantially increased, according to the data of [39] by almost three orders of magnitude, if the sample is cooled below a characteristic temperature (for the used elastomer it as about 220 K). In this case, when MAE is magnetized at low temperature, we expect to obtain magnetization curves when the particles are motionless in a magnetic field, and their locations in the MAE correspond to the original positions in the same material at room temperature in the absence of magnetic fields. The magnetization curves at low temperatures should be the same as for a composite with immobile ferromagnetic inclusions.

Thus, the scientific objective of this work is to obtain an experimental proof of the effect of matrix elasticity on the magnetic properties of an MAE. We will show that it is possible by making a comparison of magnetic properties using the data of temperature studies of the magnetic hysteresis loop.

**2. Sample preparation**

Preparation of polydimethylsiloxane-based elastomer material filled with $\varphi_m = 70$ wt% of carbonyl iron powder (CIP) was performed as reported elsewhere [23,40]. CIP (type SQ, mean diameter of particles 4.5 μm) was supplied by BASF SE Carbonyl Iron Powder & Metal Systems, Ludwigshafen, Germany. The base polymer VS 100000 (vinyl-functional polydimethylsiloxane (PDMS)) for addition-curing silicones, the chain extenders Modifier 715 (SiH-terminated PDMS), the reactive diluent polymer MV 2000 (monovinyl functional PDMS), the crosslinker 210 (dimethyl siloxane-methyl hydrogen siloxane copolymer), the Pt-catalyst 510 and the inhibitor DVS were provided by Evonik Hanse GmbH, Geesthacht, Germany. The silicone oil WACKER® AK 10 (linear, non-reactive PDMS) was purchased from Wacker Chemie AG, Burghausen, Germany.

The polymer VS 100000, the polymer MV 2000, the modifier 715 and the silicone oil AK 10 were put together and blended with an electric mixer (Roti®-Speed-stirrer, Carl Roth GmbH, Germany) to form an initial compound. In the next step, the initial compound was mixed together with the CIP particles and the crosslinker 210. The crosslinking reaction was activated by the Pt-Catalyst 510. For the control of the Pt-catalyst's activity, the inhibitor DVS was used, the recommended dosage is between 0,01 und 0,5 % [41-43]. The MAE samples were pre-cured in the universal oven Memmert UF30 (Memmert GmbH, Schwabach, Germany) at 353 K for 1 hour and then post-cured at 333 K for 24 hours with air circulation.

Two samples were prepared for the measurements. Sample 1 was a disk with a diameter of 2.0 cm and a height of 2.1 mm. Such dimensions of the sample allowed one to use a standard measurement system to determine the dynamic shear modulus. However, sample 1 has a very large magnetic moment, and its shape and large dimensions make it impossible to measure its magnetization with a vibrating sample magnetometer. To measure the magnetization using a vibration magnetometer LDJ, we employed the sample 2, which had the shape of a cylinder with a diameter of 2.5 mm and a height of 2.1 mm. Note that all properties of sample 1 and sample 2, except of their dimensions, were the same.

## 3. Magnetic field dependence of the shear modulus at room temperature

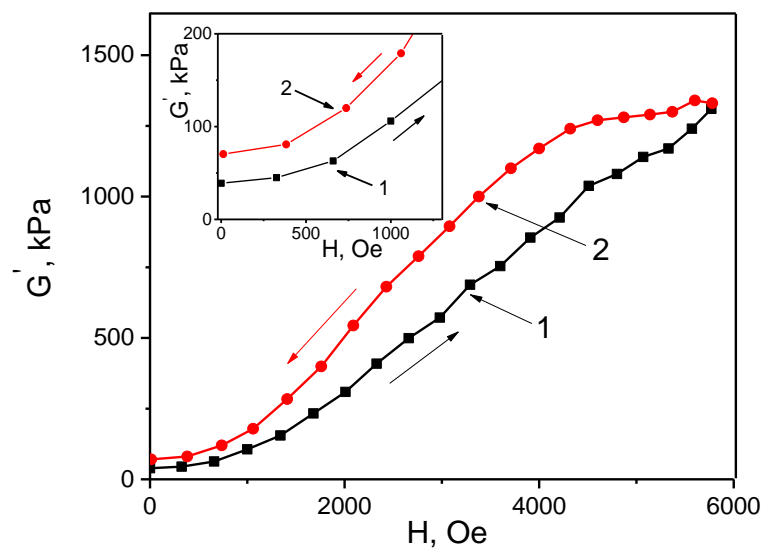

**Fig. 1.** Field dependence of the shear storage modulus $G'$ of the sample 1 for increasing (curve 1) and decreasing (curve 2) magnetic field. The inset shows the dependence in small fields.

The magnetic field dependence of the shear storage modulus $G'$ (the real part of the dynamic shear modulus) was obtained for sample 1 at room temperature and it is shown in Fig. 1. The magnitude of $G'$ is increased approximately 30-fold in the magnetic field of approximately 5.7 kOe; such a magnetorheological effect can be considered moderate at the present state of technology. In Fig. 1, the dependence of $G'$ on the increasing magnetic field strength $H$ is denoted as curve 1 (initial magnetization curve), and curve 2 denotes the dependence of $G'$ on the decreasing magnetic field strength $H$. Rheological measurements were performed using a commercially available rheometer (Anton Paar, model Physica MCR 301) with the magnetic cell MRD 170/1 T. The angular oscillation frequency $\omega$ was maintained constant at 10 rad/s. To avoid slippage, the normal force of approximately 1 N was applied. The moduli were measured at constant strain amplitude $\gamma = 0.01\%$, which corresponds to the linear viscoelastic regime. When measuring, the shape of sample 1 remained practically unchanged. The magnetic field was directed perpendicular to the surface of the disk-shaped sample and it was changed in steps. The magnitude of the field step was about 300 Oe, and the time between the steps was 20 seconds.

From Fig. 1 it is seen that the hysteresis of the $G'(H)$ dependence is observed in the entire region of the magnetization fields. Such hysteresis phenomena, including transient behavior, have been studied earlier, for example, in [44-46]. For the decreasing field $H$, the value of $G'(0)$ is larger than the value of $G'(0)$ for the initially non-magnetized MAE and the difference in these quantities turns out to be about 50% (see the inset in Fig 1.) At the subsequent ascending/descending branches of the magnetic field the value of $G'(0)$ remains almost unchanged. It can be expected that during the initial magnetization principal restructuring of the filler occurs. Initially this leads to major changes, whereas further changes are minor [23,44]. The increase of the shear storage modulus $G'$ with the growing magnetic field $H$ is commonly attributed to the additional mechanical stiffness that arises due to the interaction of magnetized particles.

## 4. Hysteresis of magnetization at room temperature

Fig. 2 shows the measured field dependence of the magnetization $m(H)$ for the sample no. 2. The sample was placed in a rigid cuvette having the same dimensions as the sample. The cuvette was not deformed during the magnetization measurements. The magnetic field was directed along the axis of the cylinder. The time between the field steps was 3 seconds, the number of points in one loop was 256, and the step magnitude for ascending and descending magnetic field $H$ was maintained constant at 155 Oe. The saturation magnetization of 150 emu/g agrees well with the expected value for a given concentration of iron particles in the sample.

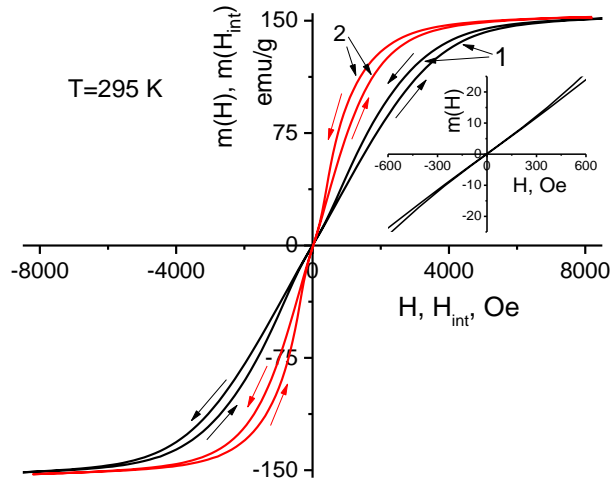

**Fig. 2.** Field dependences of the magnetization for the MAE sample. The curve 1 denotes the dependence $m(H)$, while the curve 2 denotes the dependence of the MAE magnetization $m(H_{int})$ on the internal field $H_{int}$. The inset shows the course of $m(H)$ in small fields. The arrows on both curves show the direction of the field sweep, both for its ascending and descending parts.

At $H = 0$, the remanent magnetization is absent, which agrees with the fact that the particles are made of soft magnetic iron. The ascending and descending branches of the loop coincide well only in the field interval $|H| < \pm 300$ Oe. Hysteresis regions appear at magnetic fields of larger magnitude, which is rather unusual for magnetization reversal curves. The maximum of the hysteresis width is reached for a magnetic field of roughly 2 kOe and the magnitude of the hysteresis is about 400 Oe.

The coincidence of the ascending and descending branches of the magnetic loop in small magnetic fields and the absence of the remanent magnetization can be explained that the elastic forces dominate there and they able to restore on the whole the initial configuration of soft-magnetic particles.

Fig. 2 also shows the dependence of the MAE magnetization on the internal field $H_{int}$. This dependence was obtained by eliminating the effect of the demagnetizing field, which depends on the shape of the sample. When calculating the internal field, the relative volume of the filler particles in the sample, $\varphi_V$, was taken into account. The magnitude of the demagnetizing factor of the sample was assumed to be equal to $N \approx 4\pi / 3$ (as in the case of a sphere; this is reasonable due to the shape of the sample, whose height is approximately equal to the diameter [47]). The value of the internal field was obtained using the formula

$$H_{int} = H - Nm_V \varphi_V, \quad (1)$$

where $m_V \cdot \varphi_V$ is the volume magnetization of the sample. The volume magnetization of ferromagnetic particles, $m_V$, is given by $m_V = \rho_{Fe} m(H) / \varphi_m$, where $\rho_{Fe}$ is the density of iron. As expected, the magnetization curves $m(H_{int})$ in the internal field in Fig. 2 have steeper growth and they are displaced towards the y-axis.

Let us consider the field dependences of the derivatives of the magnetization with respect to the field, that is, the field dependences for the differential magnetic susceptibility. Fig. 3 compares the field dependences of the derivatives $dm(H)/dH$ (sample susceptibility) and the magnetization derivative $dm(H_{int})/dH_{int}$ (material susceptibility), constructed according to the data shown in Fig. 2. It can be seen that, in small magnetic fields ($|H|$, $|H_{int}| < 1500$ Oe), the material susceptibility is larger than the sample susceptibility. This is explained by the demagnetization field of the sample. It is also observed that the difference between the maxima for ascending and descending magnetic fields is significantly more pronounced for the material susceptibility. We note, however, that the latter property of magnetization was absent in the Fig. 6 of Ref. 1. This can be probably explained by a significantly stiffer matrix of an MAE sample of Ref. 1. The area under the curves $dm(H)/dH$ and $dm(H_{int})/dH_{int}$) must be the same for ascending and descending branches of magnetic field.

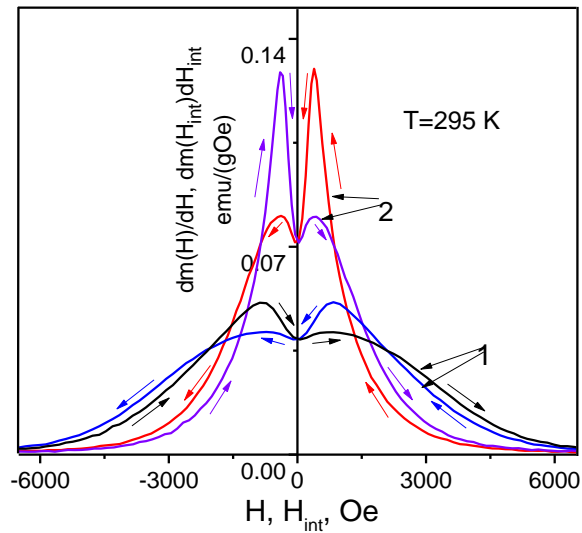

**Fig. 3.** Field dependences of differential susceptibility. Curve 1 denotes the field dependence $dm(H)/dH$, curves 2 indicates the field dependence $dm(H_{int})/dH_{int}$. The arrows designated the direction of the field sweep.

In the following when we refer to the susceptibility the differential susceptibility is implied. From the graph in Fig. 3 we obtain that the magnetization reversal of the sample has a hysteretic behavior in the entire field range. Both, when the field is increased and the field is decreased, there are local maxima in the dependence $dm(H)/dH$, and its magnitude is smaller when the absolute value of the magnetic field increases. For the decreasing absolute value of the field, the amplification of the susceptibility maximum is related to the hysteresis of the magnetization in large fields and is obviously a consequence of this hysteresis. The maxima of $dm(H)/dH$ are observed in external fields with the magnitude of 870 Oe, and the maxima of $dm(H_{int})/dH_{int}$ are observed in internal fields with the magnitude of 400 Oe.

At the point $H = 0$, the differential magnetic susceptibilities for the ascending and descending magnetic field $H$ coincide, that is, in the full loop cycle, the magnetic state at $H = 0$ is preserved.

The hysteresis of the magnetization, similar to that shown in Fig. 2, has been observed before (see, for example, [1,32,33]). Biller *et al.* [48] theoretically investigated the magnetostatic interaction of two soft-magnetic particles embedded into an elastomer matrix and showed that pair clusters of multi-domain ferromagnetic particles may form or break by a hysteresis scenario. The field dependence of the differential susceptibility, given in [1], is similar to that shown in Fig. 3. The structuring of the filler in the external magnetic field leads to the enhancement of the differential magnetic susceptibility, which in low ($H_{int} < 400$ Oe) magnetic fields exceeds the reduction of differential magnetic susceptibility due to magnetic saturation [1]. The field dependences of the relative differential permeability for iron suspensions reported in Ref. 31 have the local maximum only for the increasing magnetic field. The disappearance of the local maximum for decreasing magnetic fields has been attributed to the irreversible structuring during the first ramp of the external magnetic field [31]. However, the area below the field dependences of the relative permeability was not conserved (cf. Figs. 4 and 5 in [31]).

From a comparison of the data in Fig. 1 with the data in Fig. 2 and Fig. 3 it is seen that the magnetic and elastic properties of MAE are related, but it remains not completely clear how the elasticity of the matrix and, most importantly, how the ability of particles to change their position in magnetic field $H$ affect the magnetic properties of an MAE. It is necessary to have direct experimental proof of the effect of changing the position of the filler particles on the magnetic properties of the MAE. As we shall show below, such an experiment and the expected comparison for the magnetization of one and the same sample with moveable or "fixed" particles can be carried out with decreasing sample temperature.

## 5. Low-temperature measurements of magnetization

To investigate the intrinsic magnetization of the particles, without the influence from the elastic deformation of the matrix, the magnetization curves of the sample 2 were measured at low temperatures, the lowest temperature was 150 K. At this temperature, the elastic moduli of a homogeneous (unfilled) elastomer can increase by almost three orders of magnitude in comparison with room temperature [39]. Therefore, it can be expected that at a low temperature magnetic interactions between the inclusion particles can no longer cause noticeable deformations of the matrix and mutual displacements of the particles. As a result, the particles will be practically immobile relative to each other in a magnetic field. As the temperature of the pre-cooled sample increases, one can expect a temperature effect through the elastic softening of the matrix on the course of the magnetization reversal curve.

Fig. 4 displays the field dependences of the magnetization on external and internal magnetic fields at $T = 190$ K. There is no hysteresis of the magnetization. Practically the same hysteresis-free dependences were also observed at a lower temperature, $T = 150$ K, and at somewhat higher temperature, $T = 210$ K.

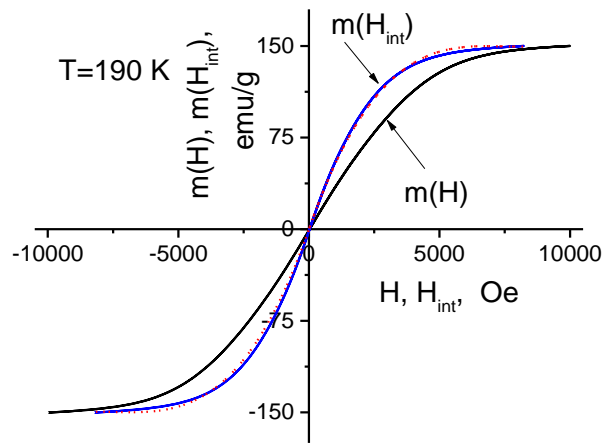

**Fig. 4.** Field dependences of the magnetization at $T = 190$ K. The curve $m(H)$ is plotted as a function of the external magnetic field strength $H$, and the curve $m(H_{int})$ is a function of the internal magnetic field strength $H_{int}$, the dashed red line denotes the fitted model.

We have fitted the experimental dependence $m(H_{int})$ shown in Fig. 4 by the following polynomial expression

$$m(H_{int}) = \chi H_{int} + R \cdot sign(H_{int})H_{int}^2 + aH_{int}^3, \qquad (2)$$

where $\chi$ is the susceptibility of a composite with fixed inclusions at $H_{int} \to 0$, parameters $R$ and $a$ describe nonlinear contributions from $H_{int}$, and sign denotes the signum function. Similar expression can be obtained from the expansion of the Fröhlich-Kenney law [49] up to the 3rd order with respect to $H_{int}$. An approximation curve was obtained for the field interval $|H_{int}| \leq 8$ kOe, its parameters are $\chi = 0.61\,\text{emu}\cdot\text{g}^{-1}\text{Oe}^{-1}$, $R = -8.22\cdot10^{-6}\,\text{emu}\cdot\text{g}^{-1}\text{Oe}^{-2}$, $a = 0.366\cdot10^{-9}\,\text{emu}\cdot\text{g}^{-1}\text{Oe}^{-3}$. The model curve (2) practically coincides with the experimental dependence $m(H_{int})$, except for a slight difference in the region where saturation is approached. The multiplier $R$ in (2) before the second-order term with respect to the magnetization can formally be called the Rayleigh constant [50]. However, it turned out that the sign of $R$ is negative. The negative sign for the second-order term in (2) is consistent with the Fröhlich-Kenney law [49]. The existence of the cubic contribution is easily verified from the second derivative of magnetization with respect to the field.

It follows from (2) that the Rayleigh constant has a negative sign, which completely corresponds to the fact that the iron particles in the MAE under investigation are soft magnetic and do not have hysteresis of intrinsic magnetization. An interpretation of the Rayleigh constant was first given by Néel, who considered the changes of magnetization brought about by the movement of domain among a series of obstacles [51]. He represented the effect of the obstacles by a "characteristic function" giving the energy of the system as a function of the position of the wall [51]. In general, the Rayleigh law of hysteresis (including the initial portion of the magnetization) should be recoverable by a macroscopic theory of hysteresis from the collective properties of interacting magnetic moments [51]. There are alternative interpretations of the Rayleigh law [51,52], but they all rely on the assumption that there are some potential barriers (obstacles, friction etc.) which have to be overcome by the magnetic field. The resulting Rayleigh constant is then positive. In our case, the Rayleigh constant is negative at low temperatures. This means that in each particle, when the magnetization is reversed, there are no potential barriers for the motion of the domain walls of the initial multi-domain state of the particle.

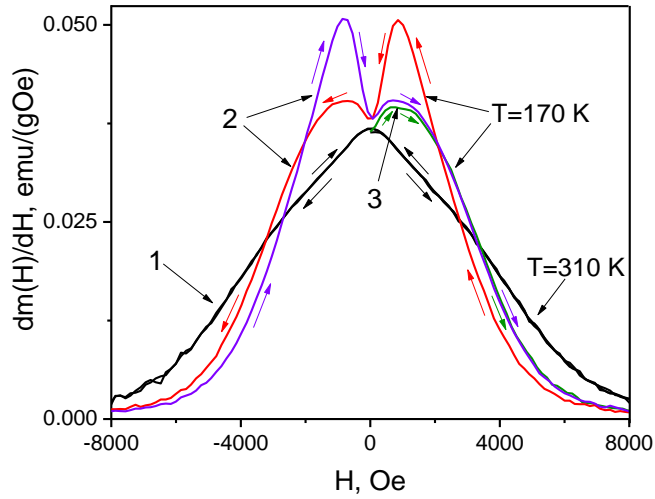

**Fig. 5.** Field dependences of the differential magnetic susceptibility at $T = 170$ K (curve 1) and $T = 310$ K (curve 2). Curve 3 designates the initial curve at $T = 170$ K.

Fig. 5 shows the field dependences of the magnetic susceptibility of the sample ($dm(H)/dH$) at $T = 170$ K (curve 1) and at $T = 310$ K (curve 2). It follows that the appearance of maxima in the field dependence of susceptibility and the magnetic hysteresis is associated with the properties of the matrix, whether it allows inclusions to change their internal positions within the MAE under an external field and deform the matrix or not.

In Fig. 5, curve 3 shows the magnetic susceptibility at $T = 310$ K, obtained from the initial magnetization curve, which was measured by increasing $H$ from 0 to 10 kOe prior to the hysteresis loop. It can be seen that the magnetic susceptibility of the initial magnetization curve at $H = 0$ and $T = 310$ K is equal to the magnetic susceptibility in $H = 0$ of the MAE with particles in the rigid matrix at $T = 170$ K, where the particles cannot change the mutual configuration during magnetization. At $T = 310$ K, the magnetic susceptibility at $H \to 0$ for the decreasing field (when the hysteresis loop is recorded) is larger than the magnetic susceptibility of the frozen sample. At high temperatures (> 220 K) in high magnetic fields (> 1000 Oe), the initial magnetization curve goes like the ascending branch of the hysteresis loop. Notice that the differential magnetic susceptibility in vanishing magnetic field is different for the initial curve and the subsequent loop. The same effect has observed for the shear storage modulus and the dielectric constant [23,44]. The explanation is that the major restructuring of the filler is completed during the initial magnetization [23,44].

Fig. 6 presents the field dependences of the magnetization normalized to the saturation magnetization $m_S$ at $T$ = 210 K, 230 K, and 310 K. The saturation magnetization $m_S$ refers to the value of $m$ in the maximal field of 10 kOe at the corresponding temperature. It is seen that at $T$ = 210 K the matrix does not affect the magnetization of the inclusions, the magnetization is without hysteresis. At $T$ = 230 K, the influence of the matrix is almost the same as at $T$ = 310 K.

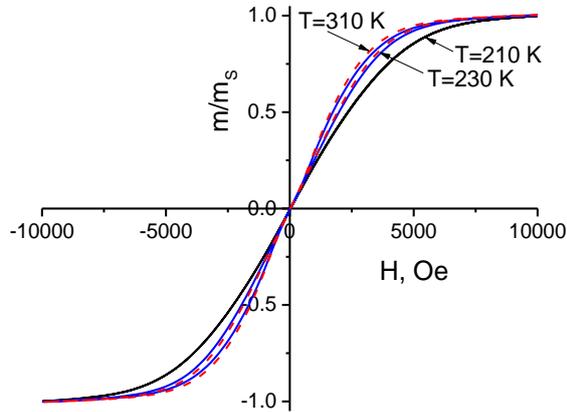

**Fig. 6**. Field dependences of the normalized magnetization $m(H)/m_S$ at temperatures $T$ = 210 K, 230 K and 310 K.

Thus, temperature changes of the course of the $m(H)$-dependence indicate that when the temperature is lowered to the 220-225 K range, the matrix "solidifies". This value of transition temperature agrees with the data of [39], obtained for an unfilled PDMS, where an important decrease of the elastic modulus upon heating was established between 203 and 233 K. It was attributed to melting of PDMS crystals. This solidification is accompanied by a sharp increase in the elastic modulus of the elastomer matrix and, as a consequence, temperature changes of the magnetization curve and the differential magnetic susceptibility confirm the effect of the mobility of the filler particles on the magnetization process of the MAE.

The appearance of a maximum in the field dependence of the differential magnetic susceptibility means that the magnetization of the particles is affected by the matrix elasticity. At low temperatures the Rayleigh constant of the filler material was found to be negative. At room temperature the differential magnetic susceptibility grows in low magnetic fields as it can be expected from the conventional Rayleigh law. An increase of temperature is unlikely to introduce additional obstacles for the movement of domain walls. We suggest that the accelerated increase in the magnetization and compensation of the negative value of the Rayleigh constant of the frozen

MAE occurs due to the dipole interaction of magnetized particles exceeding the elastic forces. The elasticity of the matrix represents the obstacle which has to be overcome by inter-particle forces which lead to restructuring of the filler.

**Conclusions**

- At low temperatures, when the rigidity of the matrix is significantly increased and the ferromagnetic filler particles, which do not possess magnetic moments in the absence of magnetic field, cannot move with respect to each other as a result of their magnetization, the magnetization of the MAE does not have hysteresis and the Rayleigh constant is negative. This corresponds to the conditions for the barrier-free motion of the particles' domain walls and "soft" magnetization reversal of the particles themselves.

- As the temperature rises, the elastomer matrix is softened in a certain temperature range (220 K - 225 K for the investigated sample). This softening is accompanied by the appearance of the magnetization hysteresis in non-vanishing fields, which shape remains practically unchanged when the MAE is heated up to the room temperature.

- The dipole interaction of particles magnetized by an external field leads to changes of their mutual configuration at high (> 225 K) temperatures, accompanied by elastic deformations of the softened matrix. As a result, the qualitative behavior of the field dependence of the differential magnetic susceptibility is changed. Its field dependence at high temperatures displays hysteresis practically in the entire region of the magnetization reversal fields. The maximum of the magnetic susceptibility is most pronounced for the decreasing absolute value of the field and is attributed to the propensity towards the restoration of the initial mutual arrangement of the filler particles caused by elastic forces during demagnetization of inclusions, which occurs when the magnetic field magnitude is decreasing.

**Acknowledgements**

V.M.K, A.A.S and M.S. thank Bayerische Forschungsallianz for financial support of the reciprocal visits (grant No. BayIntAn_OTHR_2017_120). M.S. gratefully acknowledges financial support by Deutsche Forschungsgemeinschaft (CH 1939/2-1, project No. 389008375) and OTH Regensburg (internal cluster funding).